\documentclass[11pt]{article} 
\usepackage{mystyle-new}
\usepackage{authblk}
\usepackage[T1]{fontenc}
\usepackage{epsfig,amsmath} 
\usepackage{hyperref}
\usepackage{color}
\usepackage{graphicx}
\usepackage{orcidlink}
\definecolor{red}{rgb}{1,0,0}
\def\lesssim{\ \hbox{\raise 2pt \hbox{$<$} \kern -13pt
                     \lower 3pt \hbox{$\sim$}}\ }
\def\greatersim{\ \hbox{\raise 2pt \hbox{$>$} \kern -13pt
                     \lower 3pt \hbox{$\sim$}}\ }

\def\lsim{\mathrel{\rlap{\lower4pt\hbox{\hskip1pt$\sim$}}
    \raise1pt\hbox{$<$}}}                
\def\gsim{\mathrel{\rlap{\lower4pt\hbox{\hskip1pt$\sim$}}
    \raise1pt\hbox{$>$}}}                

\def\pythia{{\sc Pythia}8}

\def\pythiatune{{\sc \textit{intrinsic}~$k_{\mathrm{T}}$\textit{-optimized}}}

\def\herwig{{\sc Herwig}}

\def\mcatnlo{{MCatNLO}}

\def\desepsf(#1 width #2){\epsfxsize=#2 \epsfbox{#1}}
\def\kt{\ensuremath{k_{\rm T}}}

\def\pt{\ensuremath{p_{\rm T}}}
\def\ptzero{\ensuremath{p_{\rm T,0}}}
\def\PZ{\ensuremath{Z}}

\def\zdyn{\ensuremath{z_{\rm dyn}}}

\def\sqrts{\ensuremath{\sqrt{s}}}
\def\ptll{\ensuremath{p_{\rm T}(\ell\ell)}}
\def\mll{\ensuremath{m(\ell\ell)}}

\newcommand{\alphas}{\ensuremath{\alpha_\mathrm{s}}}

\newcommand{\asmz}{\ensuremath{\alphas(m_{\PZ})}}

\newcommand{\mdy}{\ensuremath{m_{\small\text{DY}}}}

\newcommand{\PBM}{PB}

\newcommand{\as}{\ensuremath{\alpha_s}}

\newcommand{\GeV}{\text{GeV}}
\newcommand{\TeV}{\text{TeV}}

\newenvironment{tolerant}[1]{\par\tolerance=#1\relax}{ \par }
\usepackage{amsmath,bm}
\usepackage{lineno}

\usepackage{cite,./mcite}
\usepackage{tikz}
\usepackage[symbol]{footmisc}

\providecommand{\DOI}[1]{\href{http://dx.doi.org/#1}}

\begin{document} 

\title{
Interplay of intrinsic motion of partons and soft gluon emissions in Drell-Yan production studied with PYTHIA} 
\author[1,2]{I.~Bubanja\thanks{itana.bubanja@cern.ch}\orcidlink{0009-0005-4364-277X}}
\affil[1]{Faculty of Science and Mathematics, University of Montenegro, Podgorica, Montenegro}
\affil[2]{Interuniversity Institute for High Energies (IIHE), Universit\'e libre de Bruxelles, Belgium} 
\author[3,4]{H.~Jung \thanks{hannes.jung@desy.de}\orcidlink{0000-0002-2964-9845}}
\affil[3]{Deutsches Elektronen-Synchrotron DESY, Germany}
\affil[4]{II. Institut f\"ur Theoretische Physik, Universit\"at Hamburg,  Hamburg, Germany}
\author[1]{N.~Rai\v cevi\' c\thanks{natasar@ucg.ac.me}\orcidlink{0000-0002-2386-2290}}
\author[3]{S.~Taheri~Monfared\thanks{sara.taheri.monfared@desy.de}\orcidlink{0000-0003-2988-7859}}

\date{}
\begin{titlepage} 
\maketitle
\vspace*{-12cm}
\begin{flushright}
DESY-24-182
\end{flushright}
\vspace*{+17cm}
\end{titlepage}

\begin{abstract}
Understanding the intrinsic transverse momentum (intrinsic-\kt) of partons within colliding hadrons, typically modeled with a Gaussian distribution characterized by a specific width (the intrinsic-\kt width), has been an extremely challenging issue. This difficulty arises because event generators like \pythia\ require an intrinsic-\kt\ width that unexpectedly varies with collision energy, reaching unphysical values at high energies.
  
This paper investigates the underlying physics behind this energy dependence in \pythia , revealing that it arises from an interplay between two non-perturbative processes: the internal transverse motion of partons and non-perturbative soft gluon emissions. These contributions are most constrained in the production of Drell-Yan pairs with very low transverse momentum, where soft gluon effects become increasingly prominent with rising collision energy—contrary to initial expectations.

Through a detailed analysis of the Sudakov form factor and its influence on intrinsic-\kt\ width, we clarify the observed energy scaling behavior in \pythia , providing insight into a longstanding issue in parton shower modeling.

\end{abstract}
  
\section {Introduction}

A parton in the initial state within a hadron has longitudinal momentum in the direction of the hadron to which it belongs, 
but it also possesses a certain momentum perpendicular to the hadrons flight direction due to its internal motion (intrinsic-\kt ) inside the hadron. Studies of  intrinsic-\kt\  distributions have been recently published by several groups and collaborations~\cite{Bubanja:2023nrd, Bubanja:2024puv, CMS:2024aa}.
 While the Parton Branching (\PBM ) method~\cite{Hautmann:2017fcj,Hautmann:2017xtx}, which uses Transverse Momentum Dependent (TMD) parton distributions,  can describe the Drell-Yan (DY) pair  cross section~\cite{Martinez:2019mwt,Martinez:2020fzs} as a function of the transverse momentum,  \ptll , an intrinsic-\kt\ distribution that remains nearly independent of \sqrts\ (as shown in~\cite{Bubanja:2023nrd}) has been observed. However, in parton-shower-based Monte Carlo event generators (MC) like \pythia~\cite{Sjostrand:2014zea,Bierlich:2022pfr} and \herwig~\cite{Bellm:2015jjp}, the width of the intrinsic-\kt\ distribution depends on the center-of-mass energy, \sqrts ,  and  increases with it \cite{Gieseke:2007ad, Sjostrand:2004pf}. In Ref.~\cite{Bubanja:2024puv} it was shown that by excluding the non-perturbative region, near the soft-gluon resolution boundary in the \PBM\ method approach, a dependence of the width of the intrinsic-\kt\ distribution on \sqrts\ is observed and has similar dependence as in parton shower  MCs.

The main aim of this study is to understand the physical reason for the \sqrts -dependent width of the intrinsic-\kt\ distribution in \pythia\ 
by requiring,
that the region which is not strongly influenced by such non-perturbative contributions is well constrained, so that by varying the width of the intrinsic-\kt\ distribution, it stays stable and provides a good description of data.

This paper is organized as follows: In the first part of Section~\ref{Sec:DYsec} we discuss recent studies of the internal transverse momentum of the partons, which serve as a guide for following investigations. Additionally, we highlight the importance of the intrinsic-\kt\ on the transverse momentum of the DY pairs.   
In the second part of Section~\ref{Sec:DYsec} we discuss how the shape of the predicted DY cross section as a function of  \ptll\ changes with the choice of the selected \pythia\ tunes, focusing on the transition region between non-perturbative and perturbative parts, which is less influenced by the choice of the intrinsic-\kt\ width.
 Section~\ref{Sec:chi2} describes how  fits to DY data are performed and how  the width of the intrinsic-\kt\ distribution is determined. In section~\ref{Sec:nps}, we discuss in detail the influence of soft parton emissions on the obtained values of the width of the intrinsic-\kt\ distribution, which can be interpreted as  the interplay between  the internal transverse motion and the soft gluon emissions at very low DY transverse momenta.

\section {\label{Sec:DYsec} Low transverse momentum distribution of Drell-Yan pairs}

While the longitudinal momentum of partons inside the colliding hadrons is linked to the energy of the hadron, the internal motion of the partons inside the hadron results in a small transverse momentum (intrinsic-\kt ),  which is believed to be connected to the hadron, but independent of the hadron's longitudinal momentum.

This intrinsic-\kt\ is introduced as a non-perturbative parameter and is, for simplicity, treated as a Gaussian  distribution with a width of $\sigma$,
$e^{-k_T^2/\sigma^2}$, 
which is multiplied by the parton density function at the starting scale.

\subsection{\label{ss1} The impact of the intrinsic-\boldmath\kt\ on the transverse momentum of DY pairs }

In Ref.~\cite{Bubanja:2023nrd}, the width of the intrinsic-\kt\ distribution, $\sigma$, was determined using the \PBM\ method \cite{Martinez:2019mwt,Martinez:2020fzs} and it was shown, that $\sigma$ has little to no dependence on the center-of-mass energy, \sqrts . In the  \PBM\ method, by construction, all the soft gluon emissions were taken into account since the minimal transverse momentum of a parton emitted at any branching is close to zero. This condition is essential for obtaining collinear parton densities, which are consistent with collinear factorization.

However, this result is different from the results obtained from parton shower Monte Carlo event  generators~\cite{Gieseke:2007ad, Sjostrand:2004pf} where a center-of-mass energy dependence of the width of the intrinsic transverse momentum distribution was observed. 
In the recent paper of the CMS Collaboration~\cite{CMS:2024aa}, the  \sqrts -dependence  of the intrinsic-\kt\  width was confirmed for the event generators \pythia\ (for  tunes CP3, CP4 and CP5~\cite{Sirunyan:2019dfx})  and  \herwig\ (for tunes CH1 and CH2~\cite{CMS:2020dqt}).

In the following we show that in parton shower event generators the \sqrts -dependence  of the intrinsic-\kt\  width, $\sigma$, is coming from the suppression of small \pt\ emission by a detailed study of the low transverse momentum distributions of the DY pairs.
Next-to-leading order (NLO) calculations of inclusive DY production \footnote{Since we are focussing on small \ptll\ we do not consider higher order \as\ contribtuins, which play a role at higher \ptll\ only.} are performed using \mcatnlo~\cite{Alwall:2014hca} including a calculation of the subtraction terms relevant for the use of the \pythia\ parton shower. The final simulation is performed  with parton showers simulated by \pythia . The NNPDF3.1 NLO collinear parton densities~\cite{Ball:2017nwa} are employed for the hard process calculation. 
The hard scale of the process, $\mu$, is set to the Drell-Yan invariant mass $\mu = \mdy $. 
Predictions are compared with experimental data obtained at different collision energies, aiming to determine the best value for the width of the intrinsic-\kt\ distribution and to investigate the origin of its center-of-mass energy dependence.

The parameter that controls small \pt\ emissions in \pythia\ is the initial-state-radiation (ISR) cut-off scale,  \verb+SPACESHOWER:pT0Ref+ which has default value of 2~\GeV . 
We study the energy dependence of the width of the intrinsic-\kt\  distribution for the three values of the ISR cutoff scale: \verb+SPACESHOWER:pT0Ref+=2, 1 and 0.5~\GeV\ (in the following abbreviated as \verb+pT0Ref+).  The intrinsic-\kt\ width in the Gaussian distribution, $\sigma$, corresponds to the parameter denoted as \verb+BeamRemnants:primodialkThard+ in \pythia .  

\begin{figure} [!htb]
\includegraphics[width=0.495\linewidth]{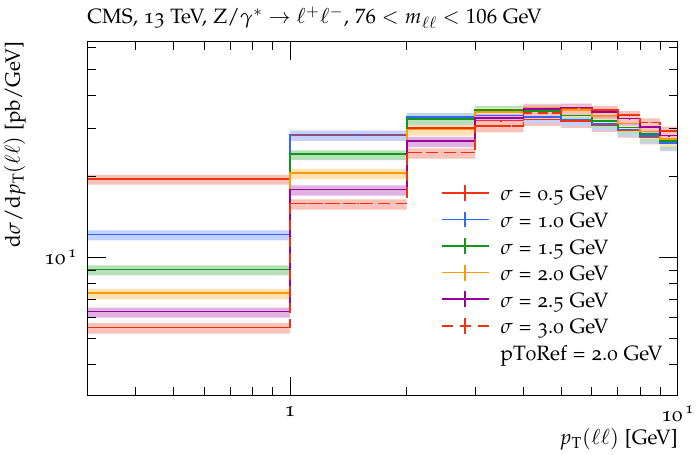} 
\includegraphics[width=0.495\linewidth]{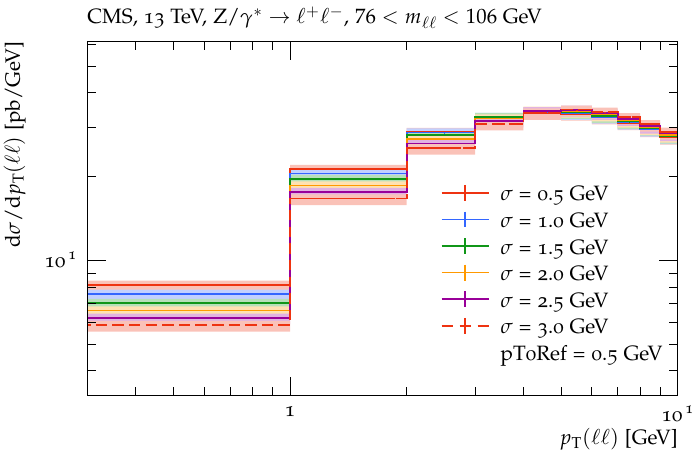}  
\caption {The DY cross section as a function of \protect\ptll\ in the \protect\PZ -peak region obtained by the default settings of \protect\pythia\ for proton-proton interactions at $\protect\sqrts = 13$~\protect\GeV\ for the two ISR cutoff parameter values, {\tt pT0Ref=2}~\protect\GeV  (left) and {\tt pT0Ref=0.5}~\protect\GeV (right), with different $\sigma$ values of $0.5, 1.0, 1.5, 2.0, 2.5$ and $3.0$~\protect\GeV . The binning follows the measurement in Ref.~\protect\cite{CMS:2022ubq}. }
\label{fig:intkt_impact}
\end{figure}

Figure \ref{fig:intkt_impact} illustrates the impact of intrinsic-\kt\ on the  transverse momentum distribution of DY pairs, \ptll , in the \PZ -peak region obtained from proton-proton collisions at 13~\TeV\ using the default settings from \pythia\  for the two values of ISR cutoff parameter: the default one, \verb+pT0Ref+=2~\GeV , and \verb+pt0Ref+=0.5~\GeV\ (all other parameters are set to the Monash 2013 tune~\cite{Skands:2014pea}). 

As shown in the figure, the impact of intrinsic-\kt\ is greatest at the lowest DY pair transverse momenta. The figure also shows that by reducing the ISR cutoff parameter, the sensitivity on intrinsic-\kt\ decreases. 
Furthermore, it indicates that the effect of the intrinsic-\kt\ distribution on \ptll\ above 3-4~\GeV\ is minimal, suggesting that this range could be used for tuning other parameters, in particular, the PDF set, the \as\ value and ordering in PDFs as well as in showers.

In the following, we illustrate how the most recent DY cross section measurements as a function of \ptll\ ,  obtained at high center-of-mass energy from LHC, $\sqrt s = 13$~TeV~\cite{CMS:2022ubq}, is described by \pythia\ using various parameter settings and tunes.

\subsection {\label{Sec:ss2} Optimal settings in \pythia }

The default setting in \pythia\ is based on the Monash 2013 tune~\cite{Skands:2014pea}, which is used to model the initial- and final-state radiation (ISR and FSR), multiparton interactions (MPI),  and hadronization. The  NNPDF2.3 LO~\cite{Ball:2012cx} parton density, with a strong coupling value of $\asmz = 0.130$, is used.  
However, a different strong coupling of $\asmz = 0.1365$ is used for ISR and FSR.  The default value of the ISR cutoff parameter is  \verb+pT0Ref+=2.0~\GeV , and the width of the  intrinsic-\kt\ distribution is set to $\sigma = 1.8$~\GeV . 

 \begin{figure} [h!]
\centering
\includegraphics[width=0.5\linewidth]{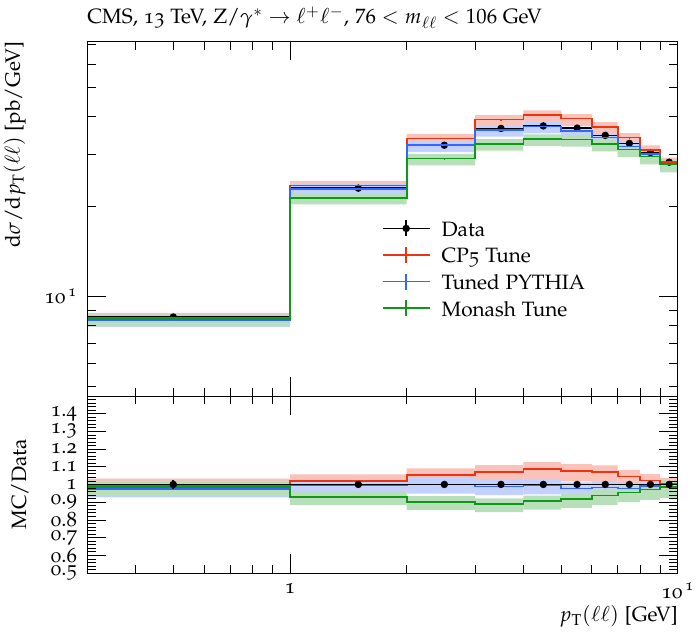} 
\caption {
Comparison of the LHC DY cross-section measurement as a function of \protect\ptll\ in the \PZ -peak region from proton-proton collisions at $\sqrts = 13$~GeV~\GeV\protect\cite{CMS:2022ubq} with \protect\pythia\ using the Monash and CP5 tunes as well as a \pythiatune\ configuration. The value {\tt pT0Ref}=0.5~\GeV\ has been used and the width of the intrinsic-\kt\ distribution has been adjusted to provide the best description of the measurement.}
\label{fig:CMS_OPT_PY_DATA_05}
\end{figure} 

The CP5 tune~\cite{Sirunyan:2019dfx} is the main tune for the underlying event (UE) in the CMS collaboration for RUN-2 data and is used for studying energy scaling behaviour in~\cite{CMS:2024aa}.
It uses a strong coupling value of $\asmz = 0.118$ and the NLO strong coupling evolution. The settings use NLO or NNLO PDFs for ISR, FSR and MPI. 
The ISR cutoff parameter is set to \verb+pT0Ref+=2.0~\GeV, and the intrinsic-\kt\ width is set to $\sigma = 1.8$~\GeV .

In Fig.~\ref{fig:CMS_OPT_PY_DATA_05}, the predictions using the Monash 2013 and CP5 tunes are compared to the measurement~\cite{CMS:2022ubq}, with \verb+pT0Ref=0.5+~\GeV\ (the minimal value accessible in \pythia ) and adjusting  the width of the intrinsic-\kt\ distribution  to best describe the cross section measurements at lowest measured \ptll .  One can observe, that using the CP5 tune leads to an overestimation of the cross section in the region of 2-5 \GeV\ while using the Monash tune results in an underestimation of the cross section in this region.
\footnote{"CMS Tune MonashStar", alias CUETP8M1-NNPDF2.3LO~\cite{Khachatryan:2015pea}, an underlying-event tune based on the Monash 2013 tune gives similar results as the default tune in \pythia . Similar results as in CP5 are obtained using CP3 and CP4 tunes~\cite{Sirunyan:2019dfx}.}
 By changing the value of $\asmz = 0.130$ in the Monash tune, an excellent description is achieved. In the following, we take this value and label it as the \pythiatune\ tune, where the intrinsic-\kt\ width is adjusted depending on $\sqrt{s}$. 

\begin{figure} [!htb]
\includegraphics[width=0.495\linewidth]{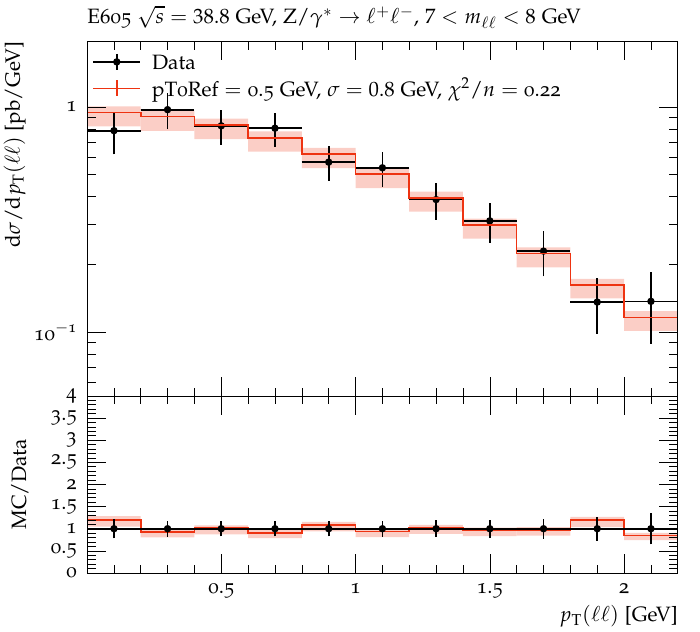} 
\includegraphics[width=0.495\linewidth]{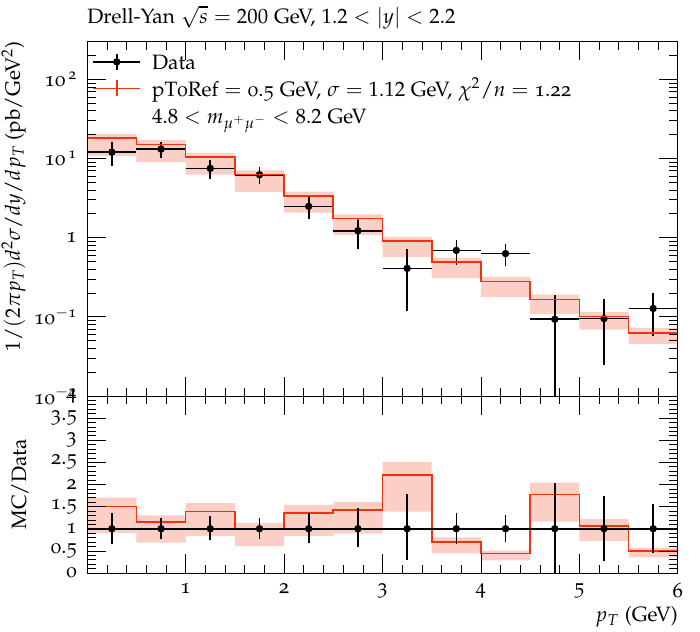}  
\caption{Comparison of the DY cross section measurement as a function of $p_{\rm T}(ll)$ in the low invariant mass region at $\sqrt{s} = 200$~GeV~\protect\cite{Aidala:2018ajl} (left) and at $\sqrt{s} = 38.8$~GeV~\protect\cite{Moreno:1990sf} (right) with \pythiatune\ tune, using {\tt pT0Ref = 0.5}~\GeV . The adjusted values of $\sigma$ are displayed.}

\label{fig:phenixe605}
\end{figure}

The \pythiatune\ tune prediction provides a very good description even at very low DY pair invariant mass when the intrinsic-\kt\ width is adjusted. 
Figure~\ref{fig:phenixe605} shows a comparison between the prediction and measurements of the DY pair cross section as a function of \ptll\ obtained from $\sqrt{s} = 200$~GeV~\protect\cite{Aidala:2018ajl}
 and $\sqrt{s} = 38.8$~GeV~\protect\cite{Moreno:1990sf} for the ranges $7 < \mll < 8$~GeV and  $4.8 < \mll  < 8.2$~GeV, respectively. The figure shows a very good agreement between the data and the prediction with a reasonable values of $\chi^2/$NDF as indicated in the figure. 
The adjusted $\sigma$ values, obtained through the procedure described in the next section, are also provided.

\section {\label{Sec:chi2} Determination of the intrinsic-\boldmath\kt\ width}  
As demonstrated in the previous section, low-energy and high-energy measurements of DY cross sections are well described using the \pythiatune\ tune with an adjusted value of $\sigma$. In this section, we outline the procedure for determining the optimal value of $\sigma$. 
The width of the intrinsic-\kt\ distribution is determined by fitting the predicted cross section to the measured DY cross section as a function of \ptll . The prediction is obtained using \mcatnlo\ supplemented with parton showers from \pythiatune\ tune. The generated events are passed through the Rivet package~\cite{Bierlich:2019rhm} for a detailed comparison with the measurement. The predictions are based on a width of the intrinsic-\kt\ distribution, which is adjusted to minimize $\chi^2$:
\begin{equation} 
\label{chi2}
 \chi^2({\tt primodialkThard}) = \sum_{i=1}^{n} \frac {{(p_i({\tt primodialkThard}) - m_i)}^2} {s^2_i}    \; ,
\end{equation}
\begin{tolerant}{8000}
where $m_i$ and $p_i$ are measured and predicted cross sections for the $i$-th bin, $n$ is the number of bins from the \ptll\ distribution used to evaluate the $\chi^2$. The value of \verb+BeamRemnants:primodialkThard+ (in the figures and formulas denoted as \verb+primodialkThard+ for brevity) for which the $\chi^2$ distribution, fitted with a cubic spline function interpolated through the points, has  a minimum is taken to be the width, $\sigma$, of the intrinsic-\kt\ contribution.  
The experimental uncertainties are treated as uncorrelated (since most published experimental data do not provide an uncertainty breakdown) and are added in quadrature to the uncertainty from the model prediction to obtain the total uncertainty $s_i$ for bin $i$.
\end{tolerant}

\begin{figure} [!h]
\includegraphics[width=0.495\linewidth]{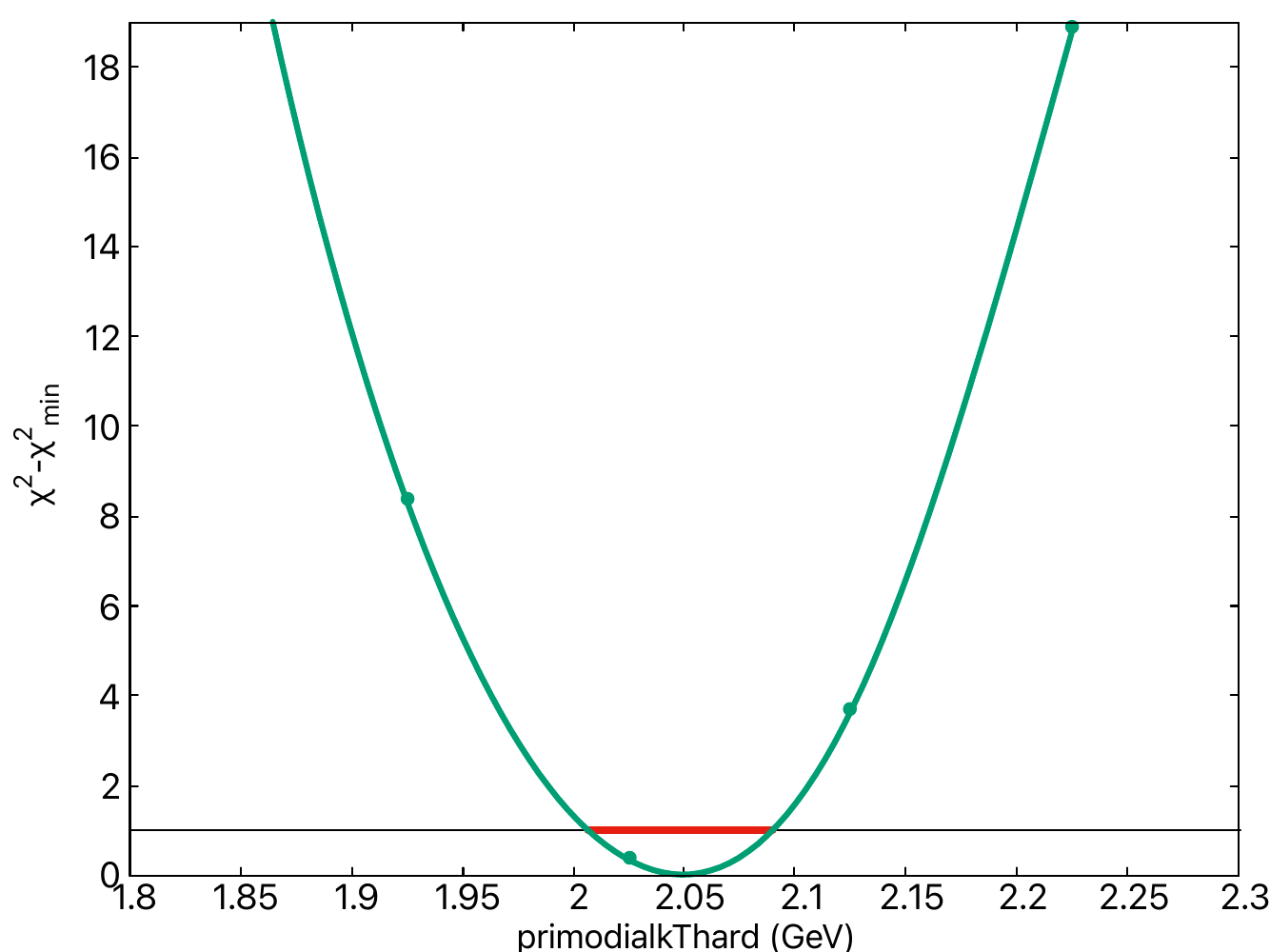} 
\includegraphics[width=0.495\linewidth]{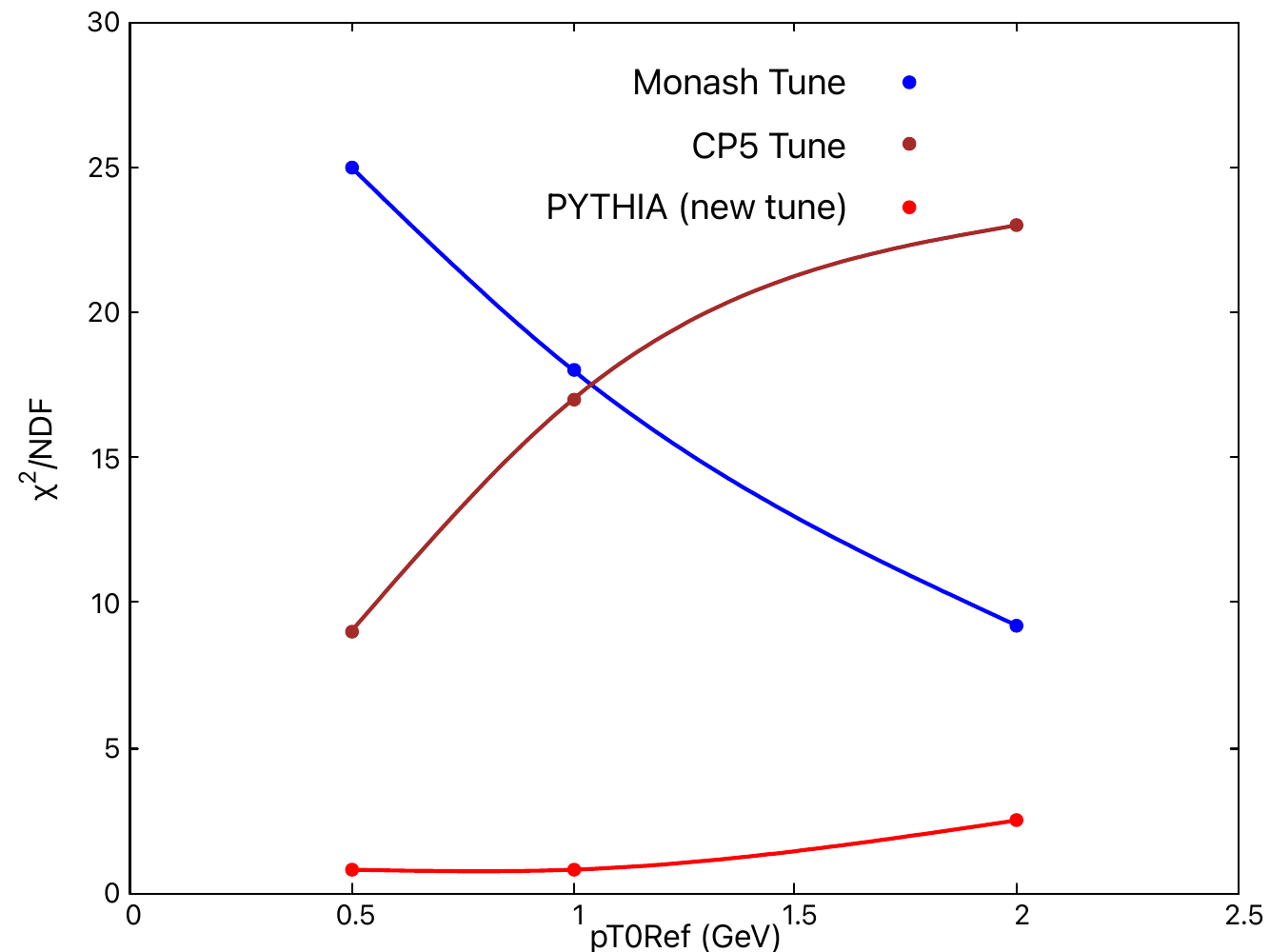} 
\caption {$\chi^2({\tt primodialkThard})-\chi^2_{\rm{min}}$ distribution obtained from the comparison of CMS  measurements at $\sqrts = 13$~\TeV~\protect\cite{CMS:2022ubq} with the \pythiatune\  tune prediction for {\tt pT0Ref} = 2~\GeV  (left). The green line represents the result of the fit using a cubic spline function interpolated through the points. The position of the minimum corresponds to the tuned parameter $\sigma$, and the red line at $\chi^2-\chi^2_{\rm min} = 1$ indicates its uncertainty. 
The right plot shows the minimum  $\chi^2/$NDF obtained in the \PZ -peak region as a function of  {\tt pT0Ref} for different tunes.
}
\label{fig:chi2}
\end{figure}

\begin{tolerant}{9000}

The uncertainty of the tuned value of $\sigma$ is estimated as the range of \verb+primodialkThard+ for which 
$\chi^2 - \chi^2_{\rm{min}}  = 1$, as illustrated by the red line in the left part of Fig.~\ref{fig:chi2}. The right part of figure~\ref{fig:chi2} shows the obtained values of $\chi^2/$NDF,  calculated from predictions obtained using the Monash, CP5 and \pythiatune\ tunes. In the following we consider only the  \pythiatune\ tune for which $\chi^2/$NDF has a reasonable value.
The number of degrees of freedom (NDF) is equal to the number of bins used in the $\chi^2$ calculation. The upper value of the  \ptll\  used for the  $\chi^2$ calculation was set to 8~\GeV\ when applicable. 
Figure~\ref{fig:chi2} shows that $\chi^2$ is significantly improved using \pythiatune\ tune compared to the other tunes. 
\end{tolerant}
\begin{figure} [ht!]
\hspace*{-0.5cm}
 \centering
\includegraphics[width=0.45\linewidth]{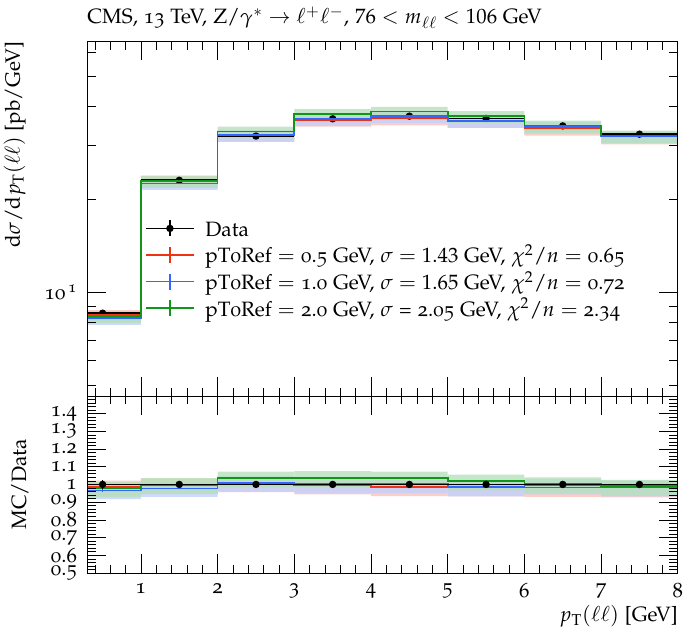}
\includegraphics[width=0.45\linewidth]{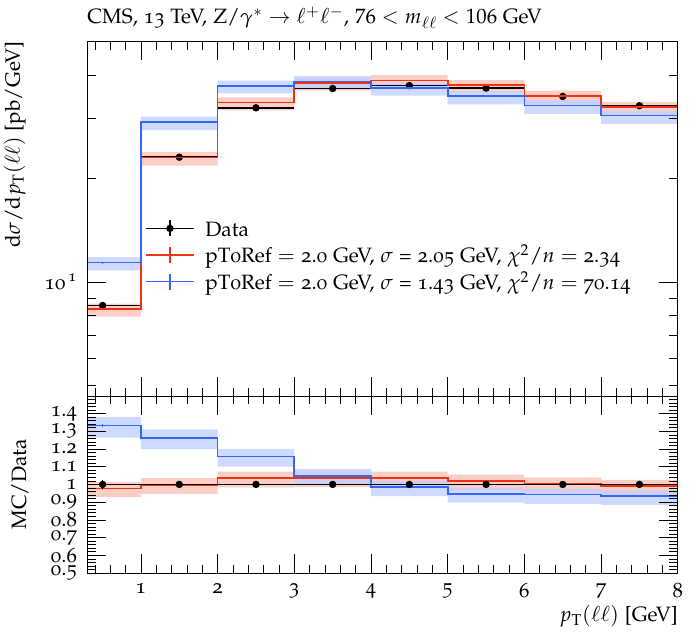} 
\caption {The cross section as a function of \protect\ptll\ in the \PZ -peak region, obtained from proton-proton collisions at $\protect\sqrts = 13$~TeV~\protect\cite{CMS:2022ubq}, compared with \protect\pythiatune\ tune. 
(Left):  Predictions using different {\tt pT0Ref} values, with the intrinsic-\kt\ width adjusted accordingly.
(Right): Prediction using {\tt pT0Ref} = 2.0 \GeV\ with both the optimal and "wrong" intrinsic-\kt\ width (the "wrong" width is the one that was obtained as an optimal one for {\tt pT0Ref} = 0.5 \GeV ).
The intrinsic-\kt\ width $\sigma$ as well as the obtained $\chi^2$ are given in the figures. }
\label{fig:CMS_20}
\end{figure}

\section {\label{Sec:nps} The impact of soft gluon emissions }

In a parton shower  language, the Sudakov form factor $\Delta (Q_1^2 , Q_2^2 ) $ is central to the splitting process, as it gives the probability of no radiation between the scales $Q_1^2$ and $Q_2^2$. 
For backward evolution in a parton shower it is:
\begin{equation}
\log \Delta_{bw} (Q_1^2, Q_2^2) = - \sum_a \int_{Q_1^2}^{Q_2^2} \frac{\as(Q^2)}{2 \pi}  \frac{d Q^2}{Q^2}\int_x^{\zdyn} dz P_{ab} (z)  \frac{x' f_b(x',Q^2)}{xf_a(x,Q^2)} \; ,
\label{Suda}
\end{equation}
with $P_{ab}$ being the DGLAP splitting function, $x =x' z$ and the scale $Q^2$ is related to the transverse momentum via  $\pt^2 = (1 -z) Q^2$ for space-like showers, leading to small \pt\ for $z$ being large. The integration limit \zdyn\ is constrained by the masses of the radiating dipole system, with $\zdyn < 1$ ~\cite{Bierlich:2022pfr}. 
The splitting probability by default is smoothly suppressed for small transverse momenta by a factor $\pt^2/(\pt^2+ \ptzero^2)$, with $\ptzero$ =  {\tt pT0Ref}. The hard scale $\mu$ of the process limits the scale $Q_2 < \mu$. 
A detailed description of the parton shower approach is given in Refs.~\cite{Bengtsson:1986gz,Bierlich:2022pfr}.
The  Sudakov form factor depends on {\tt pT0Ref} through the upper limit of the $z$ integral and on the scale $Q$ as shown in  Eq.(\ref{Suda}). 
In the study of the \PBM -method \cite{Bubanja:2024puv} a separation of perturbative and non-perturbative gluon emissions could be performed on the basis of the Sudakov form factor. In \pythia\ soft emissions are included but non-perturbaitve emissions are suppressed dynamically using  {\tt pT0Ref}, and by lowering  {\tt pT0Ref} more soft emissions are un-suppressed. It is important to note, that soft emissions can appear at all scales $Q^2$ (not only at small ones), similarly to soft emissions in the \PBM -approach. 

In the following subsections, we investigate the effect of {\tt pT0Ref}  and hard scale $\mu$ on the value of $\sigma$ for the intrinsic-\kt\ distribution, also aiming to study the contribution and distribution of soft gluons in the production of DY pairs in hadron-hadron collisions.

\subsection{\label{Sec:zdep} Dependence on \boldmath{\tt pT0Ref} }

In \pythia , the Sudakov form factor (Eq.(\ref{Suda}))  depends on the parameter \verb+pT0Ref+.
To emphasize the effect of the cutoff \verb+pT0Ref+ on the width of the intrinsic-\kt\ distribution, a comparison of the cross section measurement as a function of \ptll\ at $\sqrt s = 13$~TeV~\cite{CMS:2022ubq}, with the prediction obtained from \pythia\ is presented in 
Fig.~\ref{fig:CMS_20}.
In Fig.~\ref{fig:CMS_20}(left), we show predictions with three different values of {\tt pT0Ref}, using adjusted $\sigma$ for each  {\tt pT0Ref}. All predictions describe the measurement reasonably well, with acceptable $\chi^2$ values. 
In Fig.~\ref{fig:CMS_20}(right), we show predictions with {\tt pT0Ref} = 2.0 \GeV , using both the optimal intrinsic-\kt\ width and a  "wrong" value (the intrinsic-\kt\ width that is optimal for {\tt pT0Ref} = 0.5 \GeV ). 
One can clearly see that the value of $\sigma$ obtained for the prediction with {\tt pT0Ref} = 0.5 \GeV\ fails to describe the measurements when used with {\tt pT0Ref} = 2.0 \GeV\ since a significantly larger fraction of soft gluon contribution is removed when when the ISR cutoff parameter is increased.
This figure also  illustrates the strong influence of the soft gluon contribution with transverse momentum of the order of  1~\GeV\ at low DY pair transverse momentum.

To illustrate the effect of the cutoff \verb+pT0Ref+ on the width of the intrinsic-\kt\ distribution, we investigate several data sets for DY  production cross section as a function of \ptll ,  obtained at different center-of-mass energies: LHC data~\cite{CMS:2022ubq}, Tevatron data~\cite{CDF:2012brb}, and data from lower energies~\cite{Aidala:2018ajl, Moreno:1990sf}. The analysed data sets are shown in Table~\ref{tab:Measurements}.

\begin{table}[h!]
\centering
\begin{tabular}{|c|c|c|c|c|}
\hline
Given name & Number of bins & $E_{\rm {CM}}$ [GeV] & Ref. \\
\hline
CMS & 8 & 13000 & \cite{CMS:2022ubq} \\
CDF    & 20 & 1960 & \cite{CDF:2012brb} \\
PHENIX  & 12 & 200 & \cite{Aidala:2018ajl} \\
E605   & 11 & 38.8 & \cite{Moreno:1990sf} \\
\hline
\end{tabular}
\caption{List of measurements, collision energies, and the number of bins in \ptll\ used to determine the width of the intrinsic-\kt\ distribution.}
\label{tab:Measurements}
\end{table}

\begin{figure} [ht!]
\hspace*{-0.5cm}
  \centering
\includegraphics[width=0.9\linewidth]{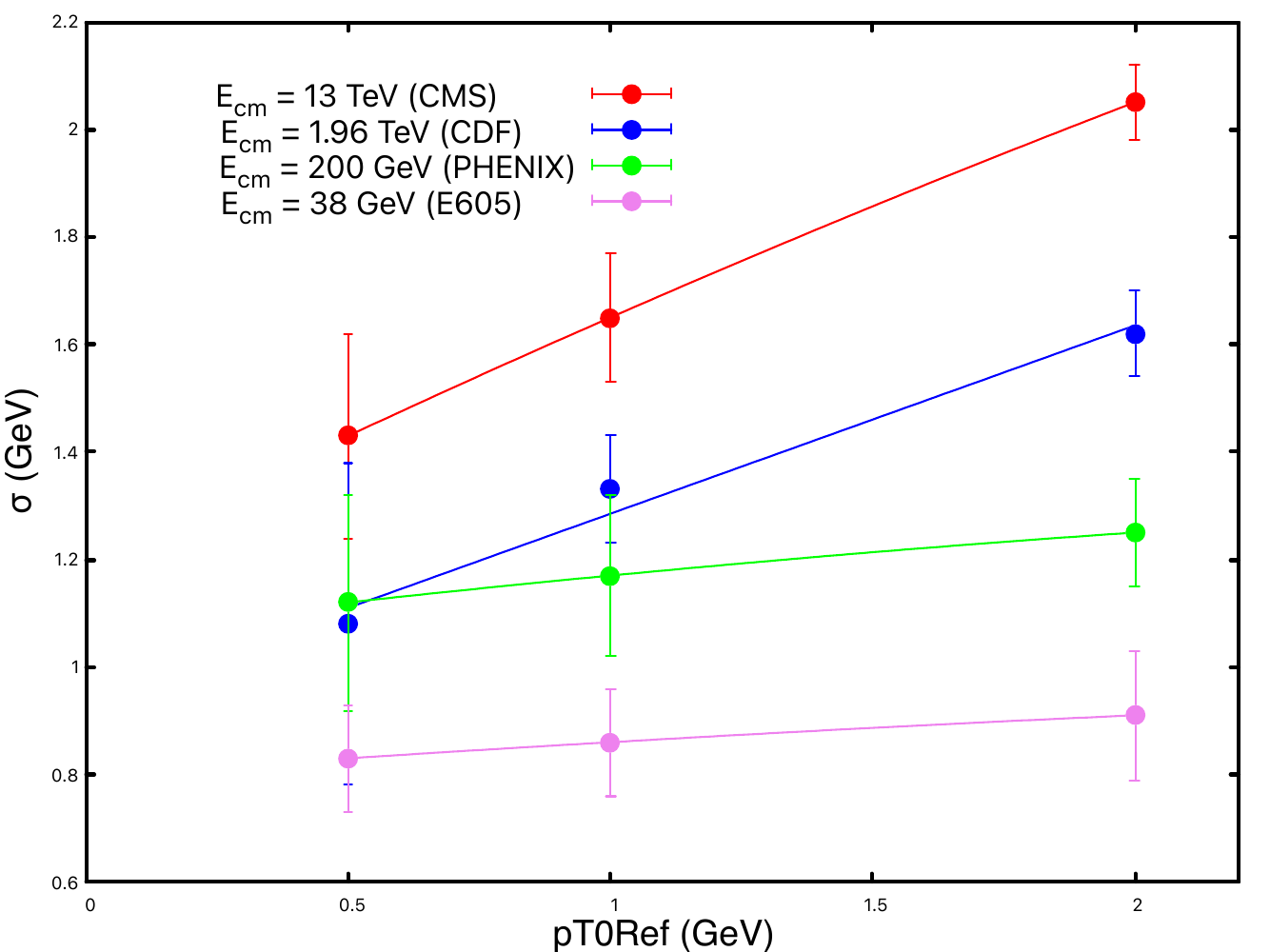} 
\caption {Dependence of the intrinsic-\kt\ width, $\sigma$, on the ISR cut-off parameter {\tt pT0Ref} used in \pythiatune\ tune,  obtained from comparison with the measurements at different collision center of mass energies listed in Table~\ref{tab:Measurements}. Linear fits are also shown for each collision energy.}  
\label{fig:qsvspt0ref}
\end{figure}

Figure~\ref{fig:qsvspt0ref} shows the width $\sigma$ obtained from $\chi^2$ minimization as a function of the  cut-off parameter, \verb+pT0Ref+, at different values of \sqrts . One can observe that for the same value of \verb+pT0Ref+, the width of the intrinsic-\kt\ distribution increases significantly with \sqrts . At higher energies, the contribution of soft, non-perturbative gluons becomes more important compared to lower \sqrts .
This observation can be understood by the increased number of branchings (emission vertices during the parton shower) at higher energies: not only does the number of perturbative emissions increase, but the number of soft, non-perturbative emissions also increases. 
In the examined region 0.5~$<${\tt pT0Ref}$<$~2.0 \GeV, the value of $\sigma$ increases nearly linearly with \verb+pT0Ref+, as indicated by the linear fits applied to the points for visual guidance. 
From the measurements at lower collision energies, 38.8 GeV~\cite{Moreno:1990sf} and 200 GeV~\cite{Aidala:2018ajl}, it is clear that the contribution from soft gluons with low transverse momentum (around 0.5 \GeV ) dominates, as the value of $\sigma$ shows only small variations with respect to the ISR cut-off parameter. 
At $\sqrts \sim \TeV$, the rate of increase of $\sigma$ with {\tt pT0Ref} is similar for $\sqrts = 13\ \TeV$ and $\sqrts = 1.96\ \TeV$.     

\subsection{\label{Sec:mudep} Dependence on the hard  scale}

From the study above, we confirm that the value of the width $\sigma$ reflects the effect of two processes: the internal parton motion and soft gluon emission. Therefore, by studying the intrinsic-\kt\ width, which quantifies the interplay between these two processes, we can investigate soft gluon emission and estimate its contribution.

\begin{figure} [h!]
\centering
\includegraphics[width=0.495\linewidth]{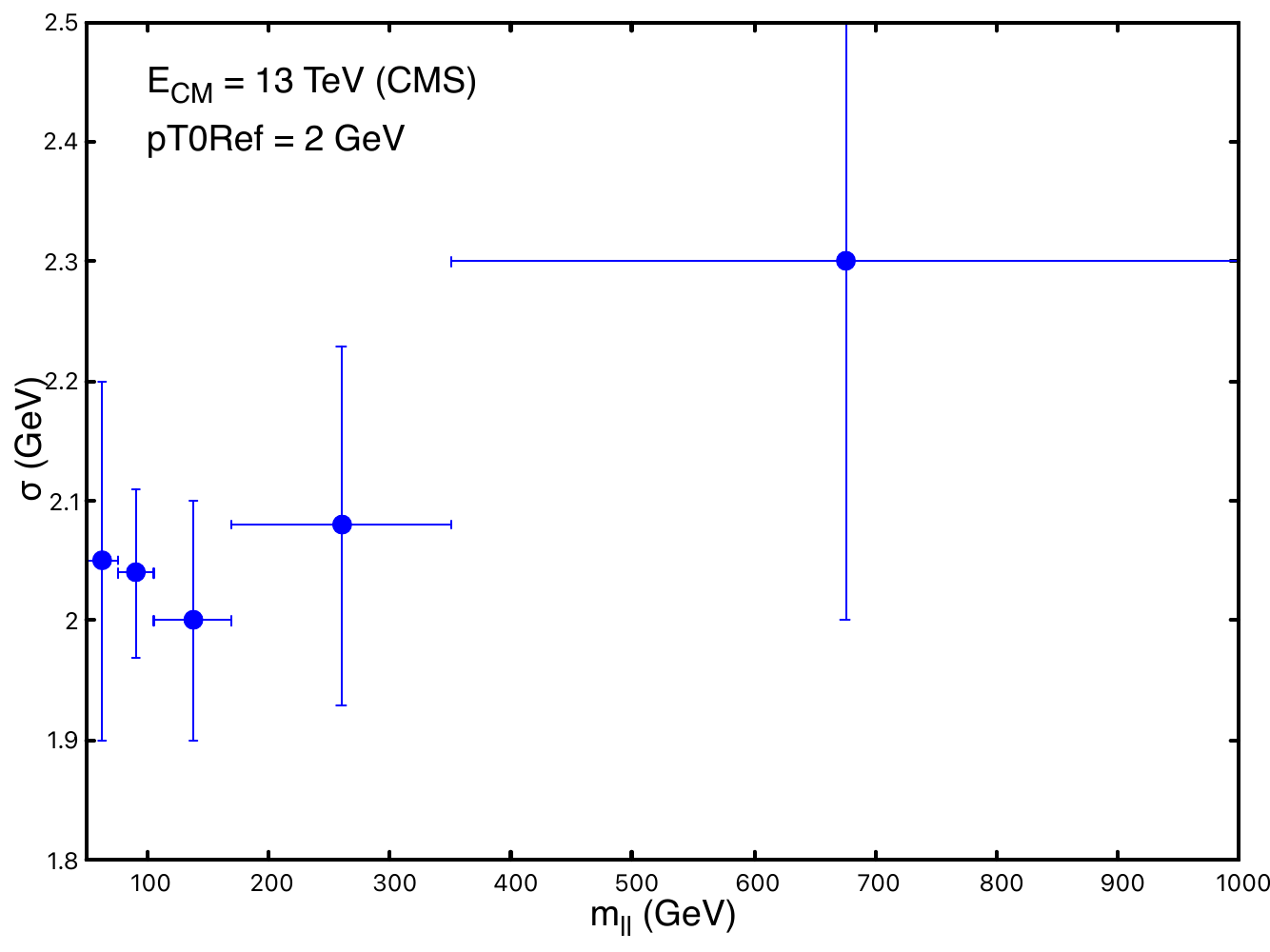} 
\includegraphics[width=0.495\linewidth]{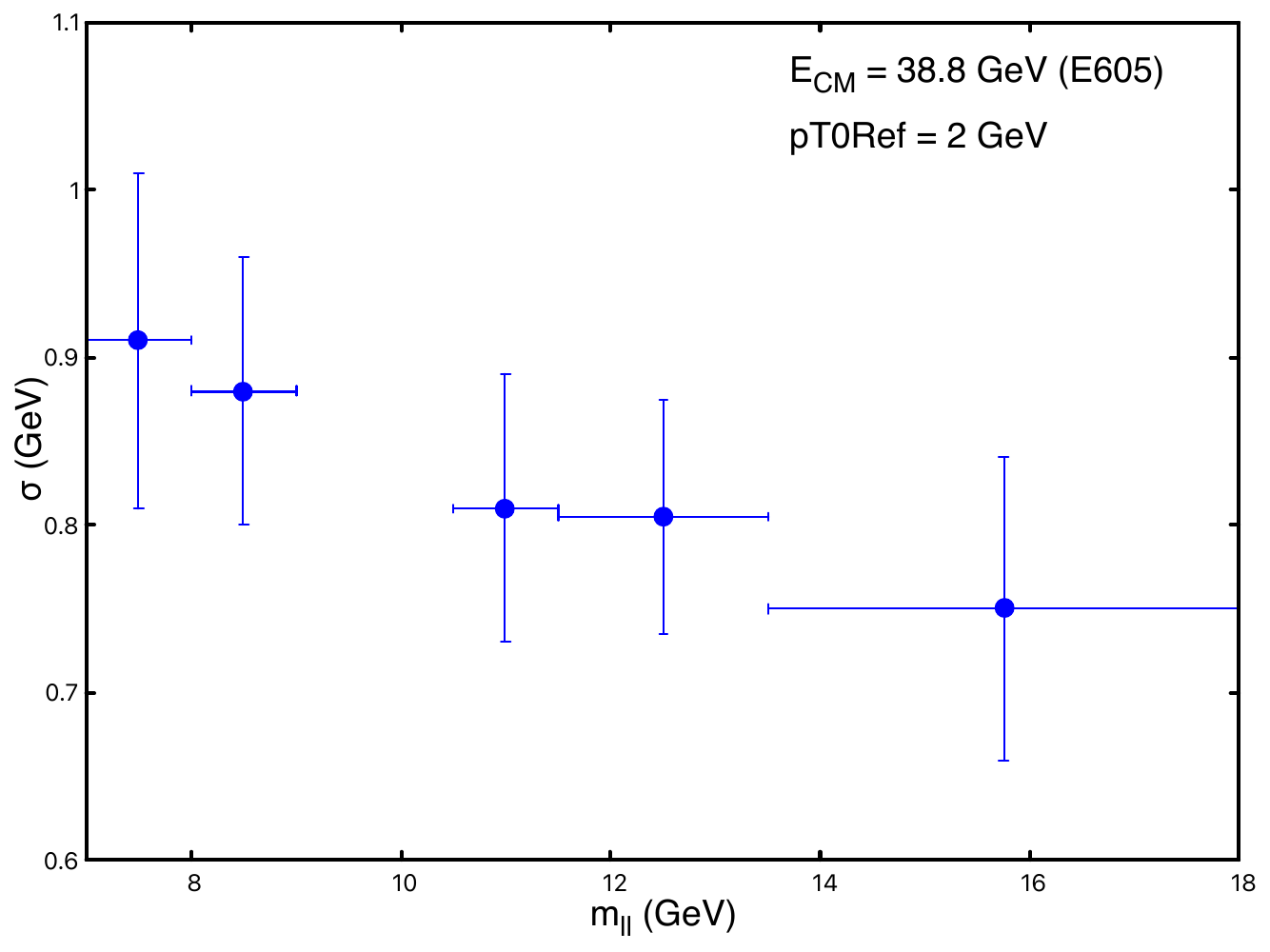} 
\caption {Intrinsic-\kt\ width, $\sigma$, as a function of the DY pair invariant mass, obtained from the comparison of data at $\sqrt s = 13$~TeV~\protect\cite{CMS:2022ubq} (left) and at $\sqrt  s = 38.8$~GeV~\protect\cite{Moreno:1990sf} (right) with \pythia\ with {\tt pT0Ref} = 2.0 \GeV . }
\label{fig:sigma_mass_dep}
\end{figure}

In order to study the impact of the scale-dependence of the soft gluon contribution, we now determine the width of the intrinsic-\kt\ distribution at low \ptll\ as a function of \mdy , which is directly related to the scale value $Q$. 
Figure~\ref{fig:sigma_mass_dep} shows $\sigma$ as a function of \mdy\ at $\sqrt{s} = 13$~\TeV ~\cite{CMS:2022ubq} (left plot) and 38.8GeV\cite{Moreno:1990sf} (right plot) for {\tt pT0Ref} = 2.0 \GeV .
The measurements at $\sqrt{s} = 13$~\TeV\ do not show a mass dependence of $\sigma$, while the measurements at the lower energy of 38.8~\GeV\ show a weak \mdy -dependence.

\begin{figure} [h!]
\includegraphics[width=0.495\linewidth]{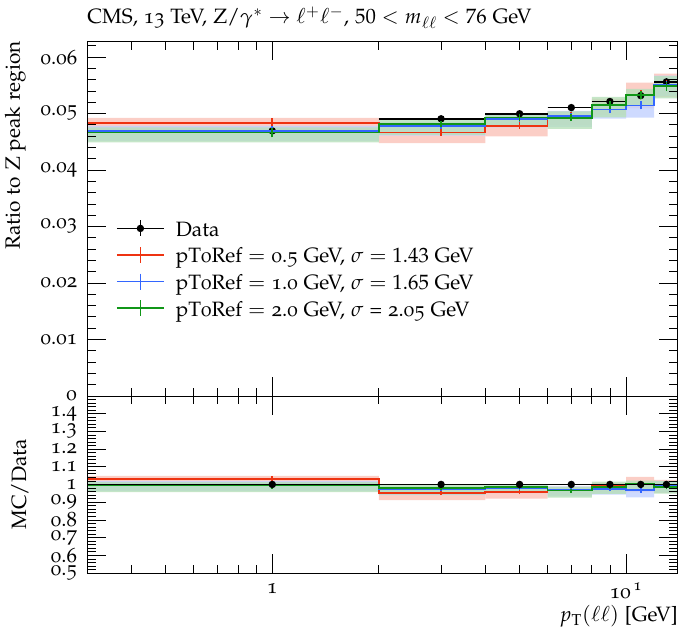} 
\includegraphics[width=0.495\linewidth]{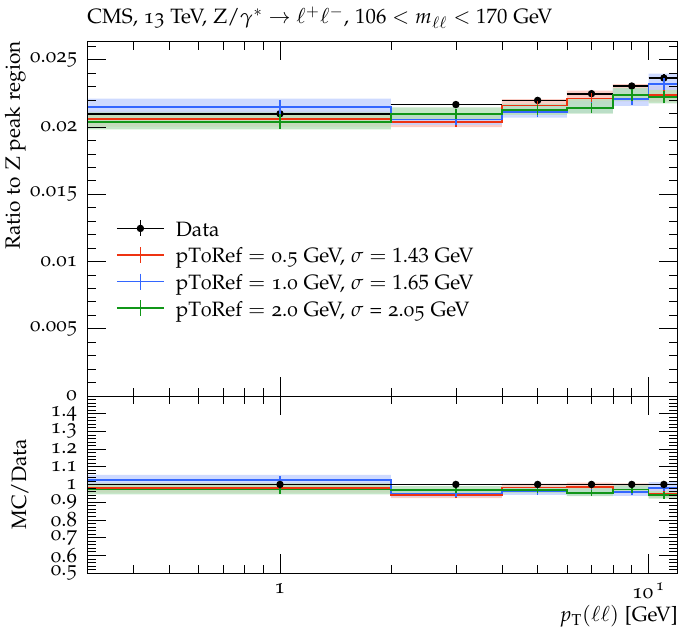} 
\includegraphics[width=0.495\linewidth]{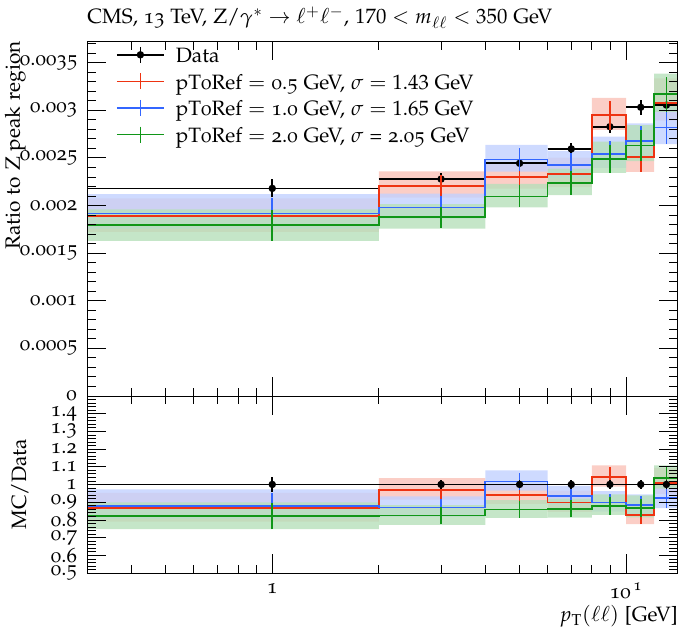} 
\includegraphics[width=0.495\linewidth]{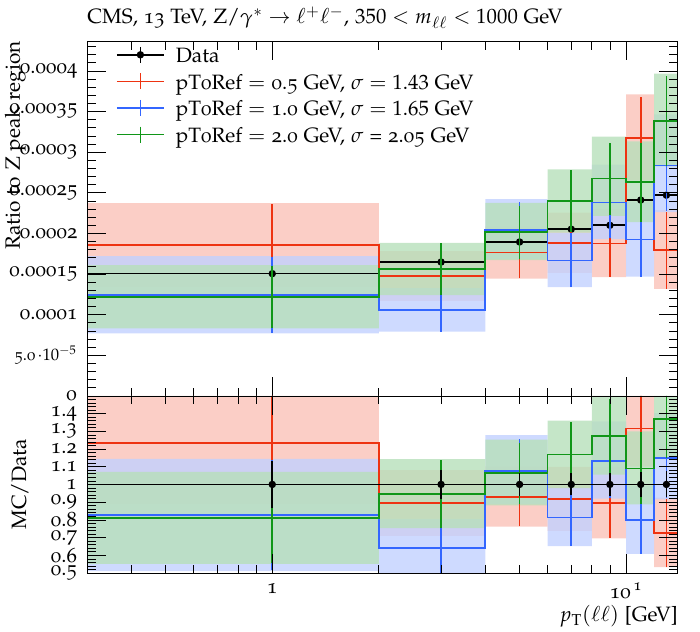} 
\caption {Measurement  of the ratio of the cross section as a function of \ptll\ to the \PZ -peak region~\protect\cite{CMS:2022ubq}, compared with predictions from \pythiatune\ tune with {\tt pT0Ref }= 0.5, 1.0 and 2.0 \GeV .}
\label{fig:Zratio_cms}
\end{figure}

The non-observation of a \mdy -dependence of  $\sigma$ suggests that the relative contribution of soft gluon emission with \pt\ below $ 2$  \GeV , which interplays with the internal transverse motion, is similar across all measured \mdy\ bins at low \ptll.

The measured ratio of the cross section as a function of \ptll\ to the one in the \PZ -peak region in several \mdy\ bins from 50 to 1000 \GeV ~\cite{CMS:2022ubq}  is shown in figure~\ref{fig:Zratio_cms} and  compared with predictions from \pythia\ for different values of {\tt pT0Ref}. All three predictions describe the measurements reasonably well, indicating that the soft gluon contribution is quite similar in all \mdy\ bins, contrary to the expectation of a scale dependence from Eq.(\ref{Suda}). 
A similar result regarding the mass dependence was obtained using the Parton Branching Method~\cite{Raicevic:2024obe}. This also agrees with the results presented in Ref.~\cite{CMS:2024aa} using the CP5 tune by the CMS Collaboration.

\begin{figure} [h!]
\centering
\includegraphics[width=0.495\linewidth]{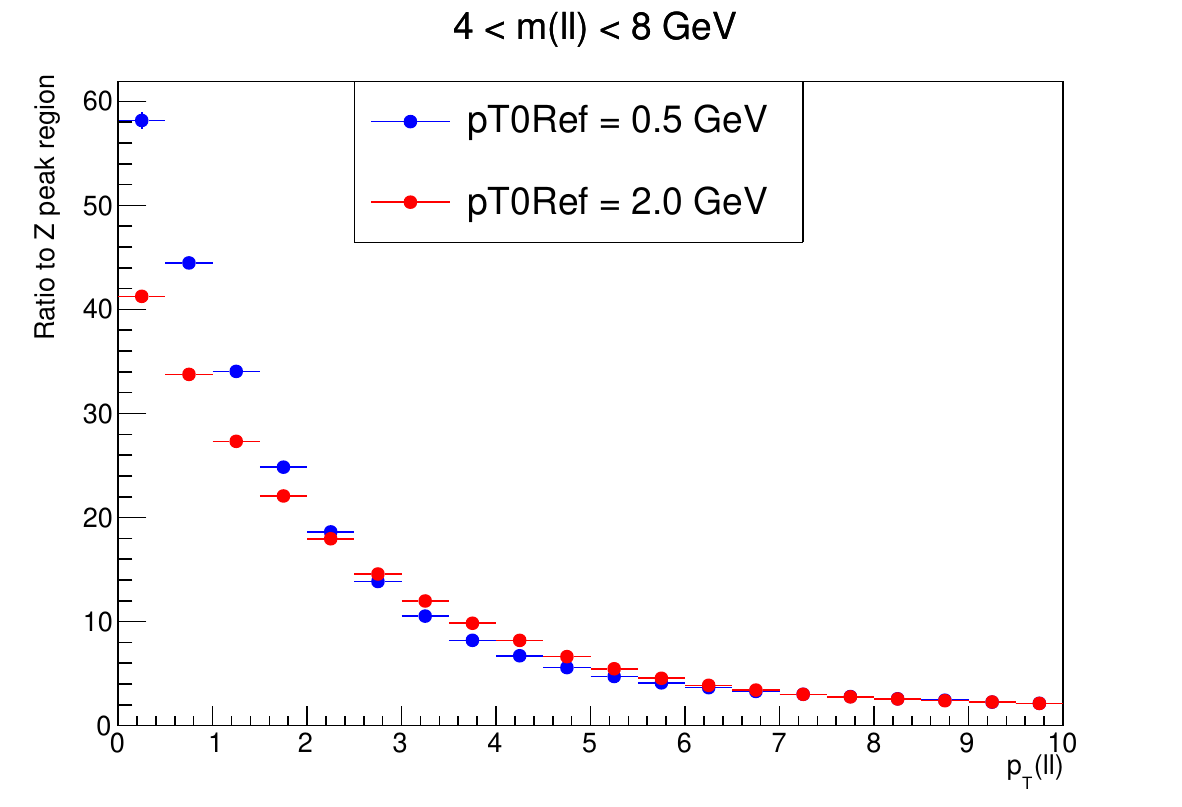} 
\includegraphics[width=0.495\linewidth]{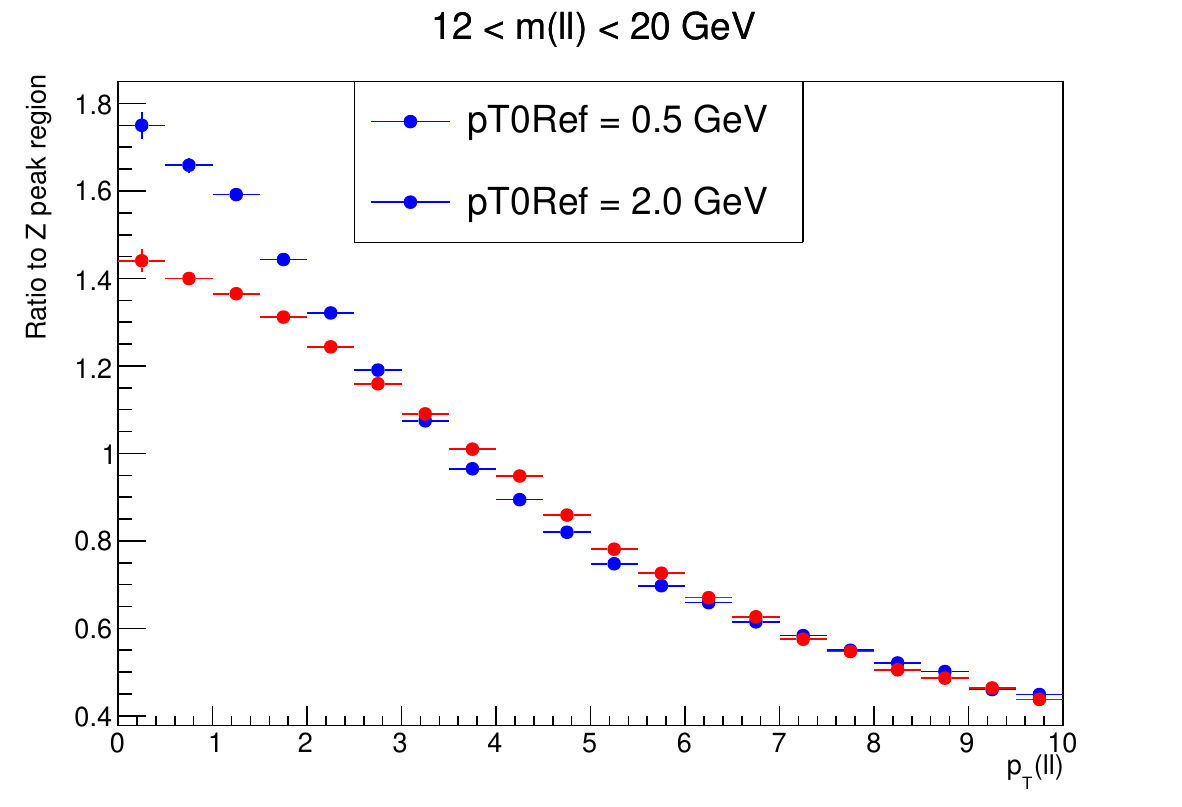} 
\includegraphics[width=0.495\linewidth]{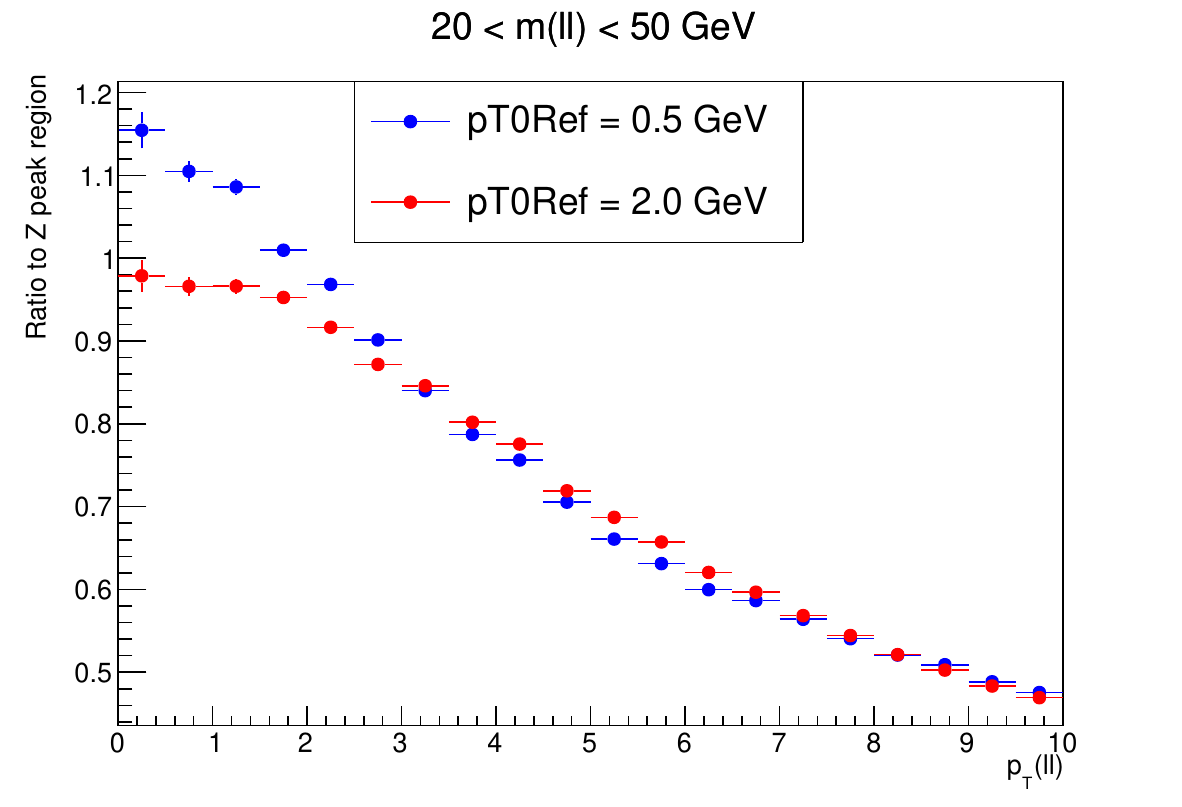} 
\includegraphics[width=0.495\linewidth]{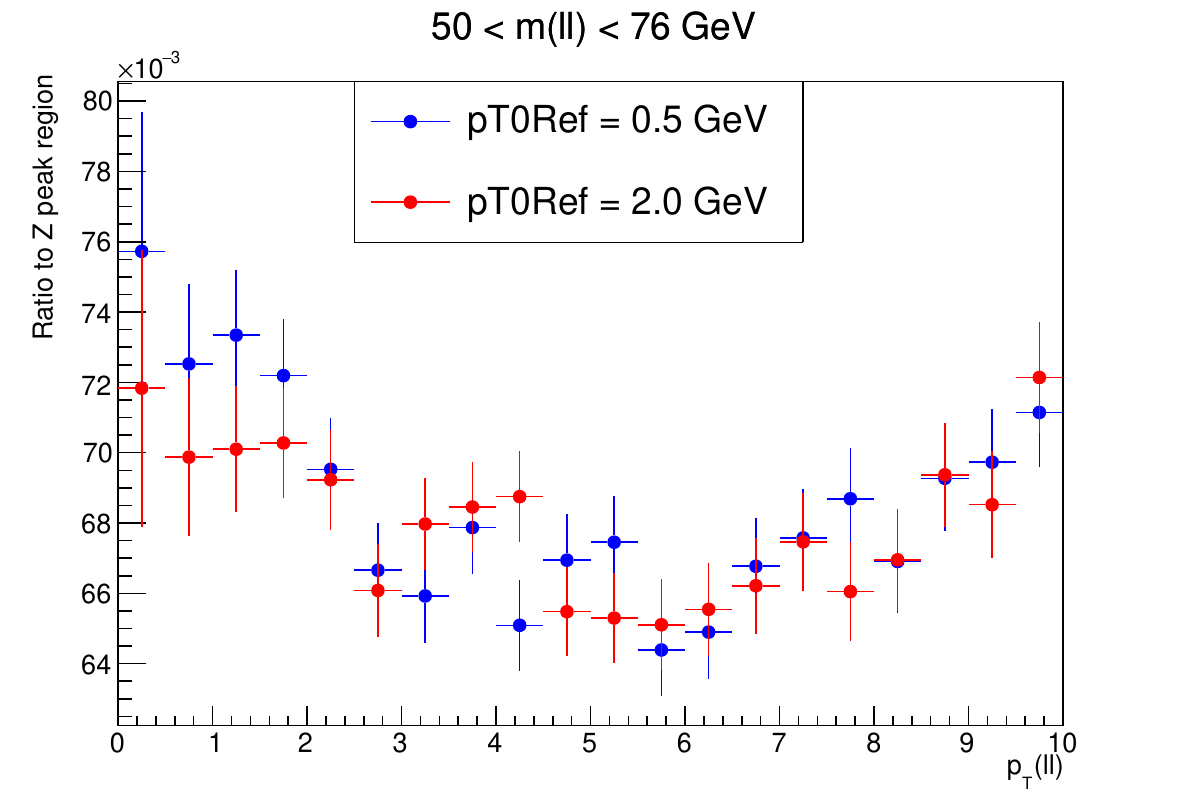}
\includegraphics[width=0.495\linewidth]{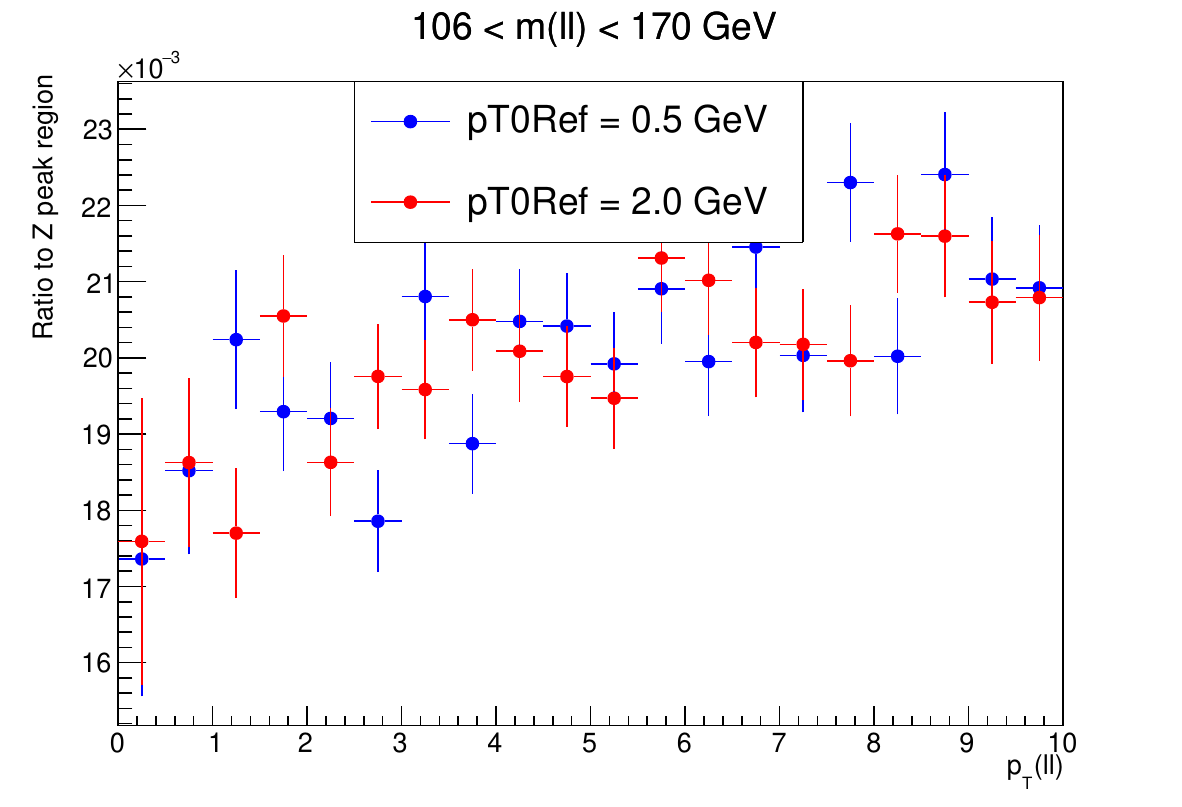} 
\includegraphics[width=0.495\linewidth]{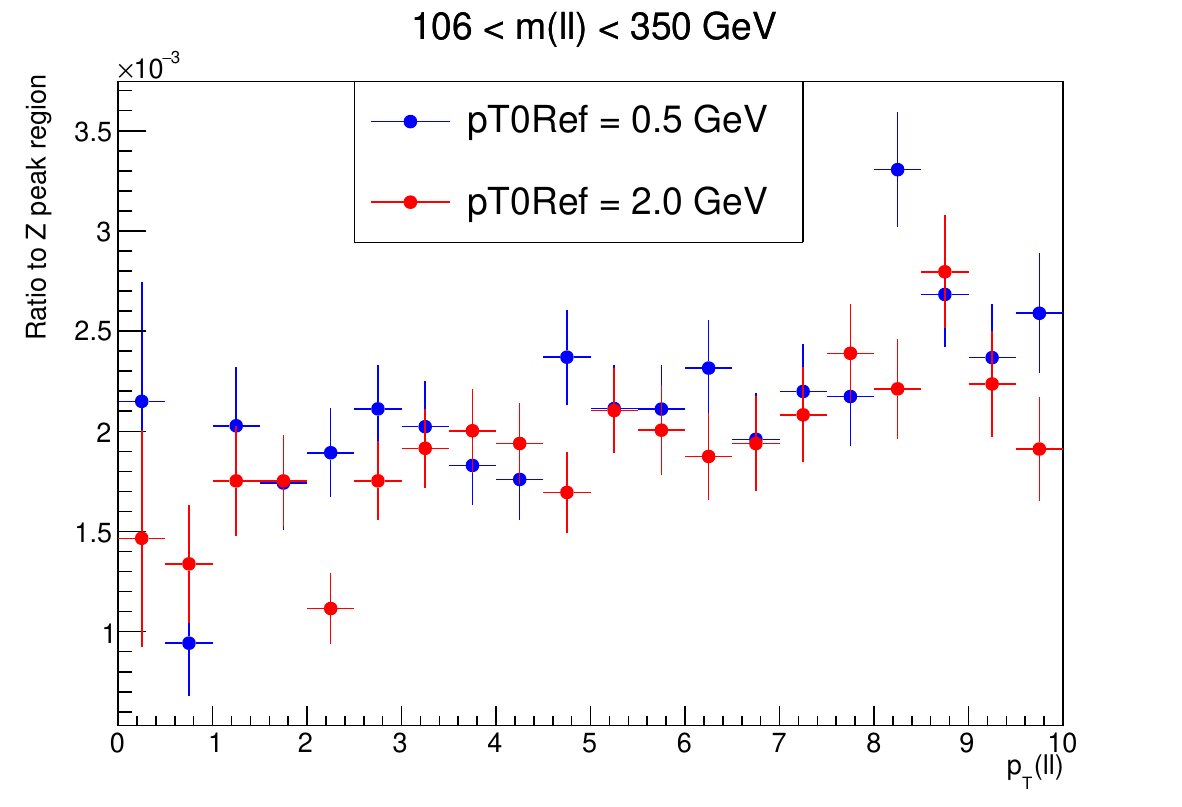} 
\caption {
The ratio of the cross section as a function of \ptll\ to \PZ -peak region from predictions from \pythiatune\ tune in different  DY pair invariant mass bins, shown for the two values of the ISR cut-off parameter: {\tt pT0Ref} = 0.5 \GeV\  and {\tt pT0Ref} = 2.0 \GeV . }
\label{fig:Zratio}
\end{figure}

To observe a scale-dependence of the width $\sigma$, we study the  ratio of the cross section as a function of  \ptll\ to that in the \PZ -peak region, using predictions from \pythiatune\ tune produced in finer binning at smaller \mdy . 

Figure~\ref{fig:Zratio} shows the ratio of the simulated cross section obtained with \pythia\ for two values of the ISR cut-off parameter: {\tt pT0Ref} = 0.5~\GeV\ and {\tt pT0Ref} = 2.0~\GeV .
For high \mdy\ values as measured at the LHC, we do not observe a significant dependence of the width parameter $\sigma$, while extending the mass range to smaller values,  such a dependence is clearly visible. The difference between the simulations with {\tt pT0Ref} = 2 \GeV\ and {\tt pT0Ref} = 0.5~\GeV\ increases as \mdy\ decreases. 
This is consistent with the expectation that most of the contribution from the soft emissions arise at low  scales.

\section {\label{Sec:sec5} Conclusion}

A detailed study was performed to accurately describe the cross section of Drell-Yan pair production as a function of transverse momentum \ptll\ using \pythia\ simulations. We focused on the region of the lowest measured transverse momentum, where non-perturbative contributions, internal transverse momentum of partons in colliding hadrons, and soft gluon emissions significantly impact the results.

We investigated the width of the intrinsic-\kt\ distribution and its relationship to soft gluon emissions to clarify the energy dependence in \pythia , as noted in several publications. We confirm that the \sqrts\ -dependence of the intrinsic-\kt\ width is connected to soft gluon emissions, which are controlled by introducing a lower cut on the transverse momentum of emitted partons in initial state radiation.

Our results demonstrate that the intrinsic-\kt\ width, reflecting the interplay between two non-perturbative processes (internal transverse motion inside the hadrons and soft gluon emissions), increases approximately linearly with the ISR cut-off parameter in the range   0.5~$< $~{\tt pT0Ref}~$<$~2.0 \GeV . This observation provides clear evidence that the \sqrts -dependence is related to the no-emission probability, Sudakov form factor, through its dependence on {\tt pT0Ref}. The $Q$-dependent part of the Sudakov form factor was also confirmed by examining the dependence of the width $\sigma$ on \mdy , which is directly linked to the evolution scale~$Q$ and is particularly visible at low \mdy .

The findings of this study address longstanding questions about the energy dependence of the intrinsic-\kt\ width in standard Monte Carlo event generators, tracing it back to contributions from soft, non-perturbative gluon emissions during parton evolution and the parton shower.

\paragraph {Acknowledgements}

We acknowledge funding of national scientific projects from the Montenegrin Ministry of Education, Science and Innovation  and the European Union's Horizon 2020 research and innovation programme under grant agreement STRONG 2020 - No 824093.  S. Taheri Monfared acknowledges the support of the German Research Foundation (DFG) under grant number 467467041.

\bibliographystyle{mybibstyle-new.bst}
\raggedright  
\providecommand{\href}[2]{#2}\begingroup\raggedright\endgroup

\end{document}